\documentclass[pra,aps,twocolumn,nofootinbib,superscriptaddress,showpacs,floatfix]{revtex4}

\usepackage{simon}
\usepackage{graphicx}
\usepackage{amsmath}

\begin{document}

\title{Alignment-to-orientation conversion and nuclear quadrupole resonance}
\author{D. Budker}\email{budker@socrates.berkeley.edu}
\affiliation{Department of Physics, University of California at
Berkeley, Berkeley, California 94720-7300} \affiliation{Nuclear
Science Division, Lawrence Berkeley National Laboratory, Berkeley,
California 94720}
\author{D. F. Kimball}\email{dfk@uclink4.berkeley.edu}
\affiliation{Department of Physics, University of California at
Berkeley, Berkeley, California 94720-7300}
\author{S. M. Rochester}\email{simonkeys@yahoo.com}
\affiliation{Department of Physics, University of California at
Berkeley, Berkeley, California 94720-7300}
\author{J. T. Urban}\email{jurban@ocf.berkeley.edu}
\affiliation{Materials Sciences Division, Lawrence Berkeley
National Laboratory, and Department of Chemistry, University of
California at Berkeley, Berkeley, California 94720-1460}

\date{\today}

\begin{abstract}
The role of alignment-to-orientation conversion (AOC) in nuclear
quadrupole resonance (NQR) is discussed. AOC is shown to be the
mechanism responsible for the appearance of macroscopic
orientation in a sample originally lacking any global
polarization. Parallels are drawn between NQR and AOC in atomic
physics.
\end{abstract}



\maketitle

The phenomenon of \emph{alignment-to-orientation conversion} (AOC)
\cite{Lom69,Hil94,Auz97,Bud2000AOC,Aln2001,Kun2002} has been of
recent interest to the atomic physics community because it is an
important physical mechanism in experiments involving the
evolution of atomic ground-state polarization\footnote{Here, we
use \emph{polarization} as a generic term describing the
anisotropy of a quantum system such as a nucleus or atom. Various
types of polarization can be described by \emph{polarization
moments} associated with corresponding spherical tensor ranks.} in
external fields. In a simple example of atomic AOC, optical
pumping by linearly polarized light produces alignment (i.e., the
rank-two quadrupole moment, which has a preferred axis but no
preferred direction) in an initially isotropic atomic ground
state. Application of a static electric field along a direction
other than that of the atomic alignment axis induces quantum beats
that result in orientation (i.e., the rank-one dipole moment,
which has angular momentum biased in one direction).

This Letter draws an analogy between the quadratic Stark splitting
encountered in atomic physics experiments and the nuclear
quadrupolar coupling encountered in solid-state nuclear
quadrupolar resonance (NQR)
\cite{Das58,Kop58,Abr62}\footnote{Studies of NQR have intensified
in recent years, with NQR in such $I = 1$ nuclei as $^{14}$N and
$^2$H finding applications in, for example, biochemistry
\cite{Edm73}, and in explosives, land mine, and narcotics
detection \cite{Yes95,Row96,Gar2001,Gre96,Gre97}.} and nuclear
magnetic resonance (NMR) experiments. In particular, we show that
in a pulsed NQR experiment, bulk magnetization is created via AOC
in an ensemble initially having zero net polarization. We present
an analytic calculation of the NQR signal produced by a powder
consisting of randomly oriented crystallites and use techniques to
visualize nuclear polarization that were developed to aid in the
understanding of polarized atomic systems. Given the similarities
between atomic physics and NQR/NMR experiments, other
opportunities may exist to use knowledge of one of these two
well-developed fields to provide insight into the other, or to aid
in the design of new experiments. For example, recent work in
atomic physics on the selective creation and detection of various
multipole moments \cite{Yas2003Select} could be adapted for use in
nuclear systems.

We first describe the relationship between the Stark and nuclear
quadrupole Hamiltonians, and then go on to discuss NQR dynamics in
more detail. The effect of an electric field $\vec{E}$, directed
along the quantization axis, on atoms of angular momentum $F$ is
described by the familiar quadratic Stark Hamiltonian:
\begin{equation}\label{eq:Stark}
    H_{E2}
    = -\frac{1}{2}\,\Ga_0 E^2
        -\frac{1}{2}\,\Ga_2 E^2\frac{3F_z^2 - F(F+1)}{F\prn{2F-1}}.
\end{equation}
Here $\Ga_0$ and $\Ga_2$ are the scalar and tensor electric
polarizabilities of the atoms, respectively. The scalar
polarizability term causes a uniform shift of all the magnetic
sublevels and therefore does not affect the ground-state
polarization dynamics. The tensor term of this Hamiltonian is
proportional to a rank-two spherical tensor operator $T_{2,0} =
3F_z^2 - F(F+1)$. A Hamiltonian consisting of only rank-one terms
acts to rotate the system, while a rank-two (or higher) term in
the Hamiltonian is generally capable of converting between
different rank polarization moments of the system.

\begin{figure}
     \center \includegraphics{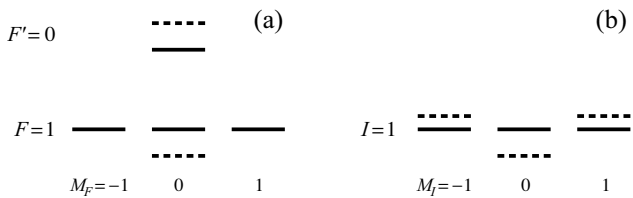} \caption{Energy
     splittings induced in (a) an $F=1$ atomic system by a uniform
     electric field, and in (b) an $I=1$ nucleus by the
     interaction of a quadrupole moment with axially symmetric
     electric field gradients. In both cases, the splitting can
     result in quantum beats that convert alignment to
     orientation.}
     \label{Fig:LevelShifts}
\end{figure}

Consider a single atomic nucleus with a nonzero quadrupole
moment\footnote{In order to possess a quadrupole moment, the
nucleus must have angular momentum $I \geq 1$. In general, a
quantum system with angular momentum $J$ may have polarization
moments of rank $\Gk$ ranging from $0$ to $2J$.} in a
polycrystalline solid. While the average electric field ``seen''
by the nucleus is zero, there may be electric field gradients that
interact with the quadrupole moment according to the
single-crystallite Hamiltonian (expressed in the Cartesian basis
$x_1=\hat{x}$, $x_2=\hat{y}$, $x_3=\hat{z}$) \cite{Abr62}
\begin{equation}\label{eq:Quad1}
    H_Q
    =-\frac{1}{6}
        \sum_{i,j} Q_{ij} \frac{\partial E_j}{\partial x_i}.
\end{equation}
Here $Q_{ij}$ is the nuclear quadrupole moment tensor and
$\vec{E}$ is the local electric field at the position of the
nucleus. Upon choosing the Cartesian coordinate system to be the
principal axis system of the local electric field gradient (EFG)
tensor, this becomes
\begin{equation}\label{eq:Quad2}
    H_Q
    =\frac{1}{3}\hbar\Go_Q
        \cbr{
            \sbr{3I_z^2 - I(I+1)}
            +\frac{\eta}{2}\prn{I_+^2 + I_-^2}
        },
\end{equation}
where
\begin{equation}
    \Go_Q
        =-\frac{3}{4I(2I-1)}
            \frac{\partial E_z}{\partial z}
            \matelem{I,m_I=I}{Q_{zz}}{I,m_I=I}
\end{equation}
is the quadrupolar sublevel splitting frequency
($\matelem{I,m_I=I}{Q_{zz}}{I,m_I=I}$ is the nuclear quadrupole
moment and $\partial E_z/\partial z$ is the principal value of the
electric field gradient tensor) and the quadrupolar asymmetry
parameter is
\begin{equation}
    \eta
    = \frac{\partial E_x/\partial x - \partial E_y/\partial y}
            {\partial E_z/\partial z}.
\end{equation}
In the following, we assume $\eta = 0$, i.e. that the electric
field gradients at each nucleus have cylindrical symmetry about
the $z$-axis. In this case, the nuclear quadrupole Hamiltonian
(Eq. \ref{eq:Quad2}) is formally analogous to the atomic Stark
Hamiltonian (Eq. \ref{eq:Stark}), with corresponding similarities
in both the level splitting (Fig.\ \ref{Fig:LevelShifts}) and the
dynamics of the two systems.

\begin{figure}
     \center \includegraphics{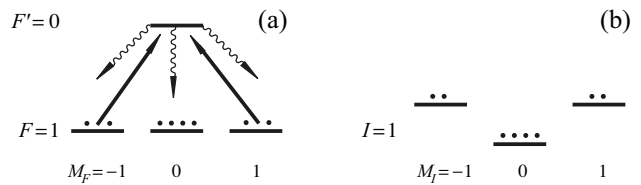} \caption{The creation of
     alignment via (a) optical pumping of an $F=1$ atomic system
     and (b) thermal distribution of quadrupole-split $I=1$
     nuclei. In (a), optical pumping with linearly polarized
     light creates atomic alignment along the polarization axis of
     the light. Straight arrows represent light-induced excitation and
     wavy arrows represent spontaneous decay. In (b), the
     alignment axis is determined by the direction of the electric
     field gradients in the crystal.}\label{Fig:OPvsThermal}
\end{figure}

The interaction (\ref{eq:Quad1}) lifts the degeneracy between
sublevels corresponding to different magnetic quantum numbers
$\abs{M}$ of the nucleus.  In a sample at thermal equilibrium, the
energy splitting gives rise to nuclear polarization because,
according to the Boltzmann law, there is a higher probability of
finding a nucleus in a lower energy state.\footnote{Typical values
of the sublevel frequency splittings are between 100 kHz and 10
MHz. At room temperature, the relative population difference
between sublevels with different $\abs{M}$ is typically
$\sim$$10^{-7}$.} Thus the nuclear quadrupolar axis is initially
along the EFG axis of symmetry. [This is in contrast to the
production of atomic alignment by optical pumping, in which the
alignment axis is determined by the polarization of the pumping
light (Fig.\ \ref{Fig:OPvsThermal}).]
Although each nucleus is in an aligned state, in a disordered
medium such as a powder there is no macroscopic polarization of
the sample because the distribution of individual crystallite
orientations is random. However, in spite of this, NQR signals
corresponding to macroscopic magnetization of the whole sample can
still be observed in such media, as discussed below.

\begin{figure}
    \center \includegraphics{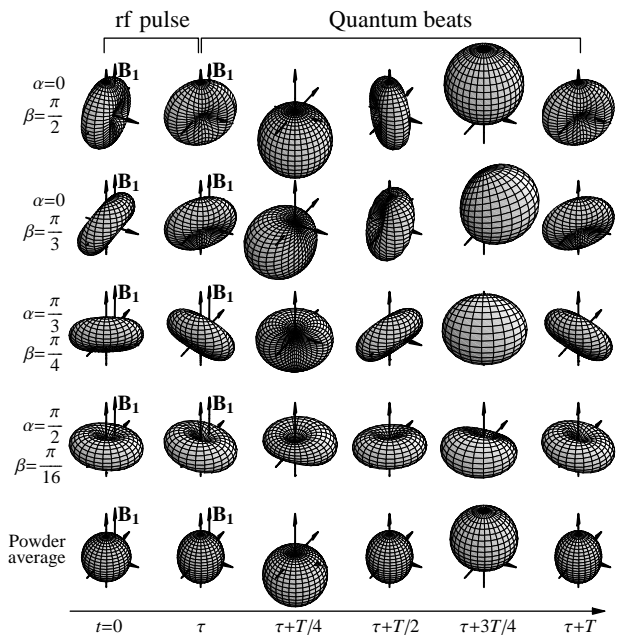} \caption{Probability surfaces
    \cite{Roc2001} corresponding to the evolution of the nuclear
    polarization in several crystallites with different
    orientations of the (axially symmetric) local field gradients.
    The angles $\alpha$ and $\beta$ are the Euler angles of the
    symmetry axes of the local electric field gradients with
    respect to the fixed lab frame. Each row shows the evolution
    of a given crystallite in time. The last row shows an average
    of the polarization over all possible crystallite
    orientations. The plots at times $t=0$ and $\Gt$ show the
    nuclear polarizations at the beginning and end of the
    excitation pulse (with magnetic field amplitude $\vec{B}_1$).
    After the pulse, a quantum beat cycle is shown. As can be seen
    in the bottom row, macroscopic oscillating orientation appears
    along the direction of $\vec{B}_1$. The plots are generated
    from a density matrix calculated using an average-Hamiltonian
    approximation in the quadrupolar interaction frame (see, for
    example, Ref.\ \cite{Lee2002}). The powder average is found by
    integrating analytically over the Euler angles (see
    Appendix).} \label{NQRFig}
\end{figure}

The initial nuclear alignment of several crystallites with
different orientations of the local field gradients is illustrated
in the first column of Fig.\ \ref{NQRFig} using angular momentum
probability surfaces, as discussed in Ref.\ \cite{Roc2001} (see
also Ref.\ \cite{Auz97}).
For a particular density matrix, the distance the surface from the
origin in a given direction is proportional to the probability of
finding the projection $M=I$ along that direction. For clarity, we
assume complete polarization, i.e., that all the nuclei are in the
lowest energy state.

The excitation is accomplished by a resonant radiofrequency (rf)
magnetic-field pulse. We assume that the field
$\vec{B}(t)=\vec{B}_1\cos(\Go t+\Gf)$ (where $\Go$ is the rf
frequency and $\Gf$ is the rf phase) has constant amplitude
$\vec{B}_1$ and is applied for a time $\Gt$ at an angle $\Gb$ to
the EFG axis of symmetry. We assume that $\Go$ is equal to the
quadrupolar frequency $\Go_Q$, and consider the decomposition of
this field into components along and perpendicular to the EFG axis
of symmetry. It can be seen that the longitudinal component causes
rapidly oscillating level shifts that have negligible effect on
the nuclear polarization, whereas the transverse component can be
further decomposed into two oppositely polarized circular
components, each of which drives one of the transitions from $M=0$
to $M'=+1$ or $-1$. Assuming that the Rabi frequency $\Go_1=\Gg
B_1$ (where $\Gg$ is the gyromagnetic ratio) is much less than
$\Go_Q$, we can neglect the nonresonant component for each
transition. The resonant components, of amplitude
$B_1\sin(\Gb)/2$, cause rotation of the nuclear polarization by an
angle $\Go_1\Gt\sin(\Gb)/2$ around the direction of the transverse
component of the magnetic field. This follows from consideration
of the dynamics in an interaction frame in which the quadrupolar
interaction is removed and the resonant components of the rf field
appear to be static (see Appendix). Since in a typical NQR
experiment the pulse length is much longer than the quantum-beat
period $T=2\Gp/\Go_Q$, quantum beats begin to occur during the rf
pulse. However, at times when the quantum-beat phase is zero, the
excitation corresponds to simple rotation. In the second column of
Fig.\ \ref{NQRFig}, we plot only the effect of the rotation, and
not of the fast quantum-beat oscillation, by assuming that the
pulse length $\Gt$ is an integer number of quantum-beat periods.
However, none of the mechanisms described here depend on this
assumption. We have chosen the parameters of the excitation pulse
such that the rotation is by $\pi/4$ for crystallites whose EFG
axes are orthogonal to the rf polarization
axis.\footnote{According to common NMR/NQR terminology, the pulse
that accomplishes such a rotation is called a $\pi/2$ pulse. The
terminology stems from the two-level spin-1/2 system, where if one
starts, for example, with a ``spin-down'' state and applies a
pulse creating a coherent superposition of ``spin-down'' and
``spin-up'' states (with equal amplitudes of the two components)
this corresponds to rotating the orientation direction by $\pi/2$.
Similarly, a $\pi$ pulse transfers all atoms from the spin-down
state to the spin-up state, and rotates the orientation by $\pi$.
In the present case of a spin-one system, if the excitation pulse
transfers all of the initial $M=0$ population into a superposition
of the $M=\pm1$ sublevels, this actually corresponds to a physical
rotation of the alignment by $\pi/2$ (not by $\pi$!).
Unfortunately, there appears to be some confusion in the
literature about this point.}

After the excitation pulse is over, the alignment axes of the
nuclei have been rotated away from the local electric field
gradient axes. Thus the nuclei are in coherent superpositions of
eigenstates of different energies---the requisite condition for
quantum beats. These quantum beats correspond to a cycle of
alignment-to-orientation conversion, as shown in the last five
columns of Fig.\ \ref{NQRFig}. In one period $T=2\Gp/\Go_Q$ of the
cycle, alignment is converted into orientation, then into
alignment at an angle of $\pi/2$ with respect to the original
alignment, followed by conversion to the opposite orientation, and
back to the original state. This evolution is the same as the
evolution of an aligned atomic system in the presence of an
electric field (see, for example, Ref.\ \cite{Roc2001}).

\begin{figure}
    \center \includegraphics{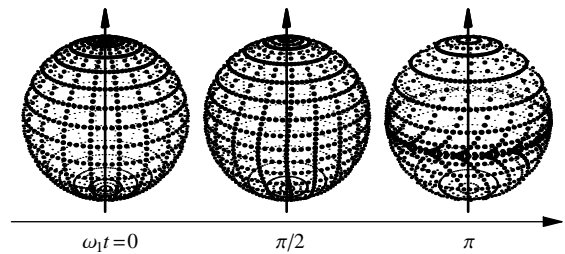} \caption{Dynamics of the
    individual crystallites' alignment axes during the excitation
    pulse. Each dot represents the intersection of a crystallite's
    alignment axis with the unit sphere. Initially ($\Go_1t=0$)
    the alignment distribution is isotropic (dots are plotted
    along parallels of latitude and meridians of longitude in
    order to illustrate the dependence of rotation on the initial
    polar angles). As the phase $\Go_1t$ accumulates, the
    alignment axes rotate (in the average-Hamiltonian
    approximation) by an angle $\Go_1t\sin(\Gb)/2$ around the
    direction of the component of the rf magnetic field transverse
    to the EFG principal axis, as described in the text. Thus, the
    alignment axes rotate away from and around the rf field axis
    (indicated by the vertical arrow).} \label{Fig:Distribution}
\end{figure}

The powder average over all crystallites is shown in the bottom
row of Fig.\ \ref{NQRFig}. Initially ($t=0$) the sample has no
average polarization, as indicated by the isotropic probability
surface. At the end of the excitation pulse of length $\Gt$ (where
$\Gt$ is chosen to be an integer multiple of $T$) there is a net
nuclear alignment within the sample, indicated by the elongation
of the probability surface along the $\vec{B}_1$ axis. The net
alignment arises because the alignment axis of each nucleus
rotates away from the $\vec{B}_1$ axis during the excitation pulse
(Fig.\ \ref{Fig:Distribution}).
Following the excitation pulse, each nucleus undergoes
alignment-to-orientation conversion. Since the orientation
produced in each crystallite is perpendicular to both the EFG
principal axis and the axis of the alignment prepared by the
excitation pulse, all crystallites contribute coherently to the
orientation along $\vec{B}_1$, which leads to a net oscillating
orientation\footnote{Even though each crystallite is undergoing
alignment-to-orientation conversion, the sample as a whole has
only oscillating orientation and no oscillating alignment. Since
the system is symmetric about the $\vec{B}_1$ axis, all
polarization not along $\vec{B}_1$ averages to zero.} of the
entire sample. This oscillating orientation corresponds to a net
sample ac magnetization that is the source of the detected NQR
signal. Details of the calculation used to obtain the
powder-averaged dynamics, including an analytic formula for the
powder-averaged density matrix, are given in the Appendix.

It must be mentioned that the conversion between polarization
moments of quadrupolar nuclei is well understood in the field of
nuclear magnetic resonance in the context of multiple-quantum
coherences. The NMR situation differs from that of NQR in several
respects. At the high magnetic field strengths common in NMR
experiments, the interaction of the nuclear spin system with this
field is dominant. Within the rotating frame approximation, the
quadrupolar interaction for any crystallite appears to be
cylindrically symmetric about the external magnetic field axis,
and the applied resonant rf fields are transverse and appear to be
static in the rotating frame. Additionally, the nuclear spin
system is initially magnetized (oriented) along the dominant
magnetic field direction. When the rf irradiation is weak compared
to the quadrupolar interaction, the conversion of orientation to
alignment has been recognized as the means by which
multiple-quantum coherence can be created from nuclear spin
magnetization during a single pulse \cite{Veg76,Veg77}. When
strong, short rf pulses are used, during which quadrupolar
evolution is negligible, the multipolar (polarization moment)
formalism has shown how multiple-quantum coherence can be created
after a two-pulse sequence via orientation-to-alignment conversion
due to quadrupolar evolution between the pulses
\cite{San83,Bow86}. Furthermore, methods have been introduced to
visualize the polarization moments of the nuclear spin system in
terms of graphical representations of the corresponding spherical
harmonics \cite{Hal84}.

Finally, we mention that various techniques for converting nuclear
alignment into orientation have been developed for the studies of
nuclear moments of short-lived nuclides \cite{Mat98,Cou99}.

In conclusion, we have shown that alignment-to-orientation
conversion plays a prominent role in the phenomenon of nuclear
quadrupole resonance, converting local nuclear alignment into
global orientation, and thus causing the appearance of a
macroscopic oscillating magnetic moment. This has been illustrated
using the method of angular momentum probability surfaces, and the
relationship of this mechanism to that of AOC in atomic physics
has been discussed. In future work, it will be interesting to
extend the present analysis to NQR and NMR in nuclei with $I>1$,
for which polarization moments higher than alignment are possible
and the transformations between these moments during quantum beats
are more complicated than alignment-to-orientation conversion
\cite{Bow86High,Roc2001}.

The authors are grateful to A. Trabesinger for the suggestion to
explore the connection between atomic physics and NQR, and to him
and S.~J.~Freedman, J.~Granwehr, A.~Pines, V.~V.~Yashchuk, and
M.~Zolotorev for useful discussions. This work has been supported
by the Office of Naval Research (grant N00014-97-1-0214), by NSF,
and by the Director, Office of Science, Office of Basic Energy
Sciences, Materials Sciences and Nuclear Science Divisions, of the
U.S. Department of Energy under contract DE-AC03-76SF00098. D.B.
also acknowledges the support of the Miller Institute for Basic
Research in Science.

\appendix

\section{Theory}

The Hamiltonian governing the NQR dynamics in the presence of an
rf field is derived, for example, in Ref.\ \cite{Lee2002}. It is
convenient to represent this Hamiltonian in the
$\prn{\ket{x},\ket{y},\ket{z}}$ Cartesian basis\footnote{Defined
by $\ket{x}=\prn{\ket{M=-1}-\ket{M=+1}}/\sqrt{2}$,
$\ket{y}=-i\prn{\ket{M=-1}+\ket{M=+1}}/\sqrt{2}$,
$\ket{z}=\ket{M=0}$.} of the EFG principal axis system (oriented
at Euler angles $\Ga$ and $\Gb$ to the rf field), because the
nuclear quadrupolar Hamiltonian (\ref{eq:Quad2}) is diagonal in
this basis. Transforming to the quadrupolar interaction frame
removes the diagonal nuclear quadrupolar Hamiltonian and adds
additional time-dependent terms in the off-diagonal elements. This
form of the Hamiltonian, given by
\begin{widetext}
\begin{equation}
    \tilde{H}(t)
    =\hbar\Go_1\cos\prn{\Go t+\Gf}
        \begin{pmatrix}
            0 &
            \cos\Gb\,e^{\frac{2i\Gh}{3}\Go_Qt} &
            \cos\Ga\sin\Gb\,e^{i\prn{1+\frac{\Gh}{3}}\Go_Qt}\\
            \cos\Gb\,e^{-\frac{2i\Gh}{3}\Go_Qt} &
            0 &
            i\sin\Ga\sin\Gb\,e^{i\prn{1-\frac{\Gh}{3}}\Go_Qt}\\
            \cos\Ga\sin\Gb\,e^{-i\prn{1+\frac{\Gh}{3}}\Go_Qt} &
            -i\sin\Ga\sin\Gb\,e^{-i\prn{1-\frac{\Gh}{3}}\Go_Qt} &
            0
        \end{pmatrix},
\end{equation}
facilitates a low-power time-average approximation that removes
rapidly oscillating terms. Assuming that the electric field
gradients are cylindrically symmetric ($\Gh=0$) and that the rf
field is resonant with the quadrupolar frequency ($\Go=\Go_Q$),
upon averaging over time-dependent terms the Hamiltonian becomes
\begin{equation}
    \bar{H}(t)
    =\frac{\hbar\Go_1}{2}
        \begin{pmatrix}
            0 &
            0 &
            \cos\Ga\sin\Gb\,e^{-i\Gf}\\
            0 &
            0 &
            i\sin\Ga\sin\Gb\,e^{-i\Gf}\\
            \cos\Ga\sin\Gb\,e^{i\Gf} &
            -i\sin\Ga\sin\Gb\,e^{i\Gf} &
            0
        \end{pmatrix}.
\end{equation}
We assume that the initial density matrix (also written in the
Cartesian basis) is fully aligned along the EFG axis:
\begin{equation}
    \Gr(0)
    =\begin{pmatrix}
        0 & 0 & 0\\
        0 & 0 & 0\\
        0 & 0 & 1
    \end{pmatrix}.
\end{equation}
The interaction-frame time dependence is given by
\begin{equation}
\begin{split}
    \tilde{\Gr}(t)
    &{}=e^{-i\bar{H}t/\hbar}\,\Gr(0)\,e^{i\bar{H}t/\hbar}\\
    &{}=\begin{pmatrix}
        \cos^2\Ga\sin^2\!\prn{\frac{1}{2}\Go'_1t}&
        -i\cos\Ga\sin\Ga\sin^2\!\prn{\frac{1}{2}\Go'_1t}&
        -\frac{i}{2}\cos\Ga\sin\!\prn{\Go'_1t}e^{-i\Gf}\\
        i\cos\Ga\sin\Ga\sin^2\!\prn{\frac{1}{2}\Go'_1t}&
        \sin^2\Ga\sin^2\!\prn{\frac{1}{2}\Go'_1t}&
        \frac{1}{2}\sin\Ga\sin\!\prn{\Go'_1t}e^{-i\Gf}\\
        \frac{i}{2}\cos\Ga\sin\!\prn{\Go'_1t}e^{i\Gf}&
        \frac{1}{2}\sin\Ga\sin\!\prn{\Go'_1t}e^{i\Gf}&
        \cos^2\!\prn{\frac{1}{2}\Go'_1t}
    \end{pmatrix},
\end{split}
\end{equation}
\end{widetext}
where $\Go'_1=\Go_1\sin\Gb$. In order to find the dynamics of the
powder average, we transform out of the interaction frame and
rotate the coordinate system to coincide with the lab frame, with
$\vec{B}_1$ along the quantization axis. The algebraic form of the
resulting density matrix is complicated. However, since in the
powder the only preferred direction is along $\vec{B}_1$, the
powder average will be symmetric about this axis. Thus, if we
represent the density matrix in the
$\prn{\ket{-1},\ket{0},\ket{1}}$ Zeeman basis, the averaged
density matrix will be diagonal, and we can ignore the
off-diagonal terms:
\begin{widetext}
\begin{equation}
    \Gr_\mr{Lab}(t)
    =\frac{1}{8}
        \begin{pmatrix}
            3-\cos\Go'_1t
                -\prn{1+\cos\Go'_1t}\cos2\Gb
                +4\sin\Go'_1t\,\sin\Gb\,\sin\prn{\Go t+\Gf}
            \ms{60}\ldots
            \ms{60}\ldots\\
            \ldots\hfill
            8\cos^2\!\prn{\frac{1}{2}\Go'_1t}\cos^2\!\Gb
            \hfill\ldots\\
            \ldots\ms{60}
            \ldots\ms{60}
            3-\cos\Go'_1t
                -\prn{1+\cos\Go'_1t}\cos2\Gb
                -4\sin\Go'_1t\,\sin\Gb\,\sin\prn{\Go t+\Gf}
        \end{pmatrix}.
\end{equation}
To obtain the result for the powder as a whole, we average over
the Euler angles, evaluating the integrals using
\emph{Mathematica} software:
\begin{equation}
\begin{split}
    \Gr_\text{avg}(t)
    &{}=\frac{
            \int_0^{2\Gp}d\Ga\int_0^{\Gp}\Gr_\mr{Lab}(t)\sin\Gb\,d\Gb
        }{
            \int_0^{2\Gp}d\Ga\int_0^{\Gp}\sin\Gb\,d\Gb
        }\\
    &{}=\frac{1}{24}
        \begin{pmatrix}
            10+4\,F'\!\prn{-\frac{1}{4}\Go_1^2t^2}
                -3\Gp\mb{H}_{-1}\!\prn{\Go_1 t}
                +8\,F''\!\prn{-\frac{1}{4}\Go_1^2t^2}\Go_1t\,\sin\prn{\Go t+\Gf}
            \ms{60}0
            \ms{60}0\\
            0\hfill
            4+4\,F'''\!\prn{-\frac{1}{4}\Go_1^2t^2}
            \hfill0\\
            0\ms{60}
            0\ms{60}
            10+4\,F'\!\prn{-\frac{1}{4}\Go_1^2t^2}
                -3\Gp\mb{H}_{-1}\!\prn{\Go_1 t}
                -8\,F''\!\prn{-\frac{1}{4}\Go_1^2t^2}\Go_1t\,\sin\prn{\Go t+\Gf}\\
        \end{pmatrix}.
\end{split}
\end{equation}
Here
\begin{equation}
    F'(z)={}_1\ms{-2}F_2\ms{-6}\sbr{\prn{2};\prn{\frac{1}{2},\frac{5}{2}};z},\quad
    F''(z)={}_1\ms{-2}F_2\ms{-6}\sbr{\prn{2};\prn{\frac{3}{2},\frac{5}{2}};z},\quad
    F'''(z)={}_1\ms{-2}F_2\ms{-6}\sbr{\prn{1};\prn{\frac{1}{2},\frac{5}{2}};z},
\end{equation}
\end{widetext}
${}_pF_{\ms{-3}q}\ms{-4}\sbr{\mb{a};\mb{b};z}$ is the generalized
hypergeometric function
\begin{equation}
    {}_pF_{\ms{-3}q}\ms{-4}\sbr{\mb{a};\mb{b};z}
    =\sum_{k=0}^{\infty}\frac{\prn{a_1}_k\ldots\prn{a_p}_k}{\prn{b_1}_k\ldots\prn{b_q}_k}\frac{z^k}{k!},
\end{equation}
where
\begin{equation}
    \prn{c}_n=c\prn{c+1}\ldots\prn{c+n-1}
\end{equation}
is the Pochhammer symbol, and $\mb{H}_n\!(z)$ is the Struve
function, which satisfies the differential equation
\begin{equation}
    z^2y''+zy'+\prn{z^2-n^2}y=\frac{2}{\Gp}\frac{z^{n+1}}{\prn{2n-1}!!},
\end{equation}
where
\begin{equation}
    n!!=n\prn{n-2}\prn{n-4}\ldots
\end{equation}
represents the double factorial.

This gives the powder average during the rf pulse. The terms
proportional to $\sin\prn{\Go t+\Gf}$ represent the quantum beats
at the quadrupolar splitting frequency. In order to find the
evolution after the pulse, we set the factors $\Go_1 t$ to a
constant corresponding to the phase accumulated during Rabi
oscillation induced by the excitation pulse. In the results
presented here, we set this constant equal to $\Gp/2$. The only
remaining time dependence is then in the $\sin\prn{\Go t+\Gf}$
terms.

\bibliographystyle{elsart-num}
\bibliography{NQR_NMR}

\begin{thebibliography}{10}
\expandafter\ifx\csname url\endcsname\relax
  \def\url#1{\texttt{#1}}\fi
\expandafter\ifx\csname urlprefix\endcsname\relax\def\urlprefix{URL }\fi

\bibitem{Lom69}
M.~Lombardi, J. Phys. (Paris) 30~(8-9) (1969) 631.

\bibitem{Hil94}
R.~Hilborn, L.~Hunter, K.~Johnson, S.~Peck, A.~Spencer, J.~Watson, Phys. Rev.
  50~(3) (1994) 2467.

\bibitem{Auz97}
M.~Auzinsh, Can. J. Phys. 75~(12) (1997) 853.

\bibitem{Bud2000AOC}
D.~Budker, D.~F. Kimball, S.~M. Rochester, V.~V. Yashchuk, Phys. Rev. Lett.
  85~(10) (2000) 2088.

\bibitem{Aln2001}
J.~Alnis, M.~Auzinsh, Phys. Rev. A 63~(2) (2001) 023407/1.

\bibitem{Kun2002}
M.~C. Kuntz, R.~Hilborn, A.~M. Spencer, Phys. Rev. A 65~(2) (2002) 023411.

\bibitem{Das58}
T.~P. Das, E.~L. Hahn, Nuclear quadrupole resonance spectroscopy, Solid state
  physics. Supplement 1, Academic, New York, 1958.

\bibitem{Kop58}
H.~Kopfermann, Nuclear moments, Academic Press, New York, 1958.

\bibitem{Abr62}
A.~Abragam, The principles of nuclear magnetism, International series of
  monographs on physics, Claredon, Oxford, 1962.

\bibitem{Edm73}
D.~T. Edmonds, C.~P. Summers, J. Mag. Reson. 12~(2) (1973) 134.

\bibitem{Yes95}
J.~P. Yesinowski, M.~L. Buess, A.~N. Garroway, M.~Ziegeweid, A.~Pines, Anal.
  Chem. 67~(13) (1995) 2256.

\bibitem{Row96}
M.~D. Rowe, J.~A.~S. Smith, Mine detection by nuclear quadrupole resonance, in:
  EUREL International Conference. The Detection of Abandoned Land Mines: A
  Humanitarian Imperative Seeking a Technical Solution, IEE, Conf. Publ.No.431,
  London, UK, 1996, p.~62.

\bibitem{Gar2001}
A.~N. Garroway, M.~L. Buess, J.~B. Miller, B.~H. Suits, A.~D. Hibbs, G.~A.
  Barrall, R.~Matthews, L.~J. Burnett, IEEE Transactions on Geoscience and
  Remote Sensing 39~(6) (2001) 1108.

\bibitem{Gre96}
V.~S. Grechishkin, A.~A. Shpilevoi, Phys. Usp. 39~(7) (1996) 713.

\bibitem{Gre97}
V.~S. Grechishkin, N.~Y. Sinyavski\u{i}, Phys. Usp. 40~(4) (1997) 393.

\bibitem{Yas2003Select}
V.~V. Yashchuk, D.~Budker, W.~Gawlik, D.~F. Kimball, Y.~P. Malakyan, S.~M.
  Rochester, physics/0302079, to appear in Phys. Rev. Lett. (2003).

\bibitem{Roc2001}
S.~M. Rochester, D.~Budker, Am. J. Phys. 69~(4) (2001) 450.

\bibitem{Lee2002}
Y.~K. Lee, Concepts in Magnetic Resonance 14~(3) (2002) 155.

\bibitem{Veg76}
S.~Vega, T.~W. Shattuck, A.~Pines, Phys. Rev. Lett. 37~(1) (1976) 43.

\bibitem{Veg77}
S.~Vega, A.~Pines, J. Chem. Phys. 66~(12) (1977) 5624.

\bibitem{San83}
B.~C. Sanctuary, T.~K. Halstead, P.~A. Osment, Mol. Phys. 49~(4) (1983) 753.

\bibitem{Bow86}
G.~J. Bowden, W.~D. Hutchison, J. Mag. Reson. 67~(3) (1986) 403.

\bibitem{Hal84}
T.~K. Halstead, P.~A. Osment, B.~C. Sanctuary, J. Mag. Reson. 60~(3) (1984)
  382.

\bibitem{Mat98}
K.~Matsuta, T.~Minamisono, Y.~Nojiri, M.~Fukuda, T.~Onishi, K.~Minamisono,
  Nucl. Instrum. Methods A 402 (1998) 229.

\bibitem{Cou99}
N.~Coulier, G.~Neyens, S.~Teughels, D.~L. Balabanski, R.~Coussement,
  G.~Georgiev, S.~Ternier, K.~Vyvey, W.~F. Rogers, Phys. Rev. C 59~(4) (1999)
  1935.

\bibitem{Bow86High}
G.~J. Bowden, W.~D. Hutchison, J.~Khachan, J. Mag. Reson. 67~(3) (1986) 415.

\end{thebibliography}







\end{document}